\documentclass[10pt,letterpaper]{article}
\usepackage{opex3}

\usepackage{amsmath}
\usepackage{mathtools}
\usepackage{amssymb}
\usepackage{graphicx}
\usepackage{epstopdf}
\usepackage{color}
\usepackage{cancel}
\usepackage{opex3}
\usepackage{cite}
\usepackage{subfigure}

\begin{document}

\title{High quantum-efficiency photon-number-resolving detector for photonic on-chip information processing}

\author{Brice Calkins$^{1,*}$, Paolo L. Mennea$^3$, Adriana E. Lita$^1$, Benjamin J. Metcalf$^2$, W. Steven Kolthammer$^2$, Ant\'ia Lamas Linares$^1$, Justin B. Spring$^2$, Peter C. Humphreys$^2$, Richard P. Mirin$^1$, James C. Gates$^3$, Peter G. R. Smith$^3$, Ian A. Walmsley$^2$, Thomas Gerrits$^1$,  and Sae Woo Nam$^1$}
\address{
	$^1$National Institute of Standards and Technology, \\* Boulder, CO, 80305, USA \\
	$^2$Clarendon Laboratory, University of Oxford, Parks Road, \\* Oxford, UK, OX1 3PU, United Kingdom\\
	$^3$Optoelectronics Research Centre, University of Southampton, \\* Highfield SO17 1BJ, United Kingdom\\
}
\email{$^*$brice.calkins@nist.gov}

\date{\today}

\begin{abstract*}
The integrated optical circuit is a promising architecture for the realization of complex quantum optical states and information networks.  One element that is required for many of these applications is a high-efficiency photon detector capable of photon-number discrimination.  We present an integrated photonic system in the telecom band at 1550 nm based on UV-written silica-on-silicon waveguides and modified transition-edge sensors capable of number resolution and over 40~\% efficiency.  Exploiting the mode transmission failure of these devices, we multiplex three detectors in series to demonstrate a combined 79~\% $\pm$ 2~\% detection efficiency with a single pass, and 88~\% $\pm$ 3~\% at the operating wavelength of an on-chip terminal reflection grating.  Furthermore, our optical measurements clearly demonstrate no significant unexplained loss in this system due to scattering or reflections.  This waveguide and detector design therefore allows the placement of number-resolving single-photon detectors of predictable efficiency at arbitrary locations within a photonic circuit -- a capability that offers great potential for many quantum optical applications.

*Contribution of NIST, an agency of the U.S. government, not subject to copyright\\
\end{abstract*}


\bibliographystyle{osajnl}
\bibliography{onchip_refs}

\begin{thebibliography}{10}
\bibitem{note:Wconductivity}
Value is based on experimentally measured low-temperature normal resistance of
  $\approx 10~\Omega$ for a 20 nm thick square device. Note that this is for a
  W film that consists of a mixture of crystallographic phases and is not
  typical for bulk W.
\bibitem{note:TESnoise}
We assume a typical simple model for the noise spectrum of the device as per
  Ref \cite{IrwinHilton}. We find experimentally that this model fits
  reasonably well, although a more sophisticated approach will be necessary to
  describe the noise of these devices exactly.
\end{thebibliography}


\section{Introduction}
Integrated optics has emerged as a leading technology to construct and manipulate complex quantum states of light, as desired for diverse applications in quantum information and communication. In contrast to bulk devices, photonic integrated circuits allow large-scale interferometric networks with excellent mode-matching, inherent phase-stability, and small footprints. Whilst quantum integrated photonic experiments have already shown on-chip two-photon quantum interference~\cite{Politi2008, Smith2009}, a current challenge is to further increase this complexity~\cite{Metcalf2012, Spring2013}. This increase will require improved total device efficiencies – spanning the generation, manipulation, and detection of single photons. A promising route is to engineer monolithic devices in which each of these functions occur on a single substrate, thus avoiding interface losses. To this end, a variety of approaches to on-chip photon sources~\cite{Uren2004, Eckstein2011} and reconfigurable circuits~\cite{Smith2009, Shadbolt2011} are being pursued. 

The first on-chip single-photon detectors have been recently realized by fabricating detectors that absorb the evanescent field of a guided wave. Adapting leading cryogenic superconducting detectors that had previously been used only as normal-incidence absorbers, evanescently-coupled on-chip detection has been achieved with a niobium nitride superconducting nanowire single-photon detector (SNSPD)~\cite{Sprengers2011, HongTang2012} and a tungsten transition edge sensor (TES)~\cite{Gerrits2011}. Where SNSPDs provide exceptionally low timing jitter and a short recovery time~\cite{Hadfield:2009}, TESs allow photon-number resolution with negligible dark counts~\cite{LitaAE08}. In addition to avoiding losses from coupling off chip, these evanescently-coupled photon detectors offer a number of benefits compared to normal-incidence devices, including efficient detection over a wide range of wavelengths, polarization-dependent detection, arbitrary placement of a detector within a planar circuit, and transmission of the undetected signal.  This final point means that a detector of low efficiency $\eta$ is equivalent to a beamsplitter-detector system with a splitting ratio of $(1 - \eta ) / \eta$ with a detector of 100~\% efficiency on the $\eta$-coupling arm.  This capability is of great significance to quantum optics applications involving heralding or photon subtraction~\cite{Gerrits:2010, ourjoumtsev:qc2006b,neergaard-nielsen:qc2006a,wakui:qc2007a,takahashi:qc2008a}.

In order to realise the full potential of these on-chip detectors for use in more complex quantum photonic circuits they must replicate the high quantum efficiencies achieved by their bulk counterparts. In this paper we report an advance in the achievable quantum efficiency of an on-chip TES detector. By carefully designing the device geometry accounting for both the optical and thermal properties of the device we have achieved  43~\% on-chip quantum efficiency.  Taking advantage of the intrinsic photon-number resolution provided by the TES together with the fact that any undetected signal is transmitted along the waveguide, we have demonstrated that it is possible to achieve close to unity quantum efficiency by multiplexing several detectors along the guide. Using three detectors placed one after another, a total quantum efficiency of 79~\% $\pm$ 2~\% is realized. The addition of an integrated Bragg reflector after the final device increases this total detection efficiency to 88~\% $\pm$ 3~\%. These results are also in agreement with both theoretical optical modeling and room-temperature characterization of the device, which confirms the absence of significant scattering losses in the system and allows the performance of future detectors to be accurately assessed and optimized before use.


\section{Device design}

\begin{figure}
\centering
\includegraphics[width=0.9\textwidth]{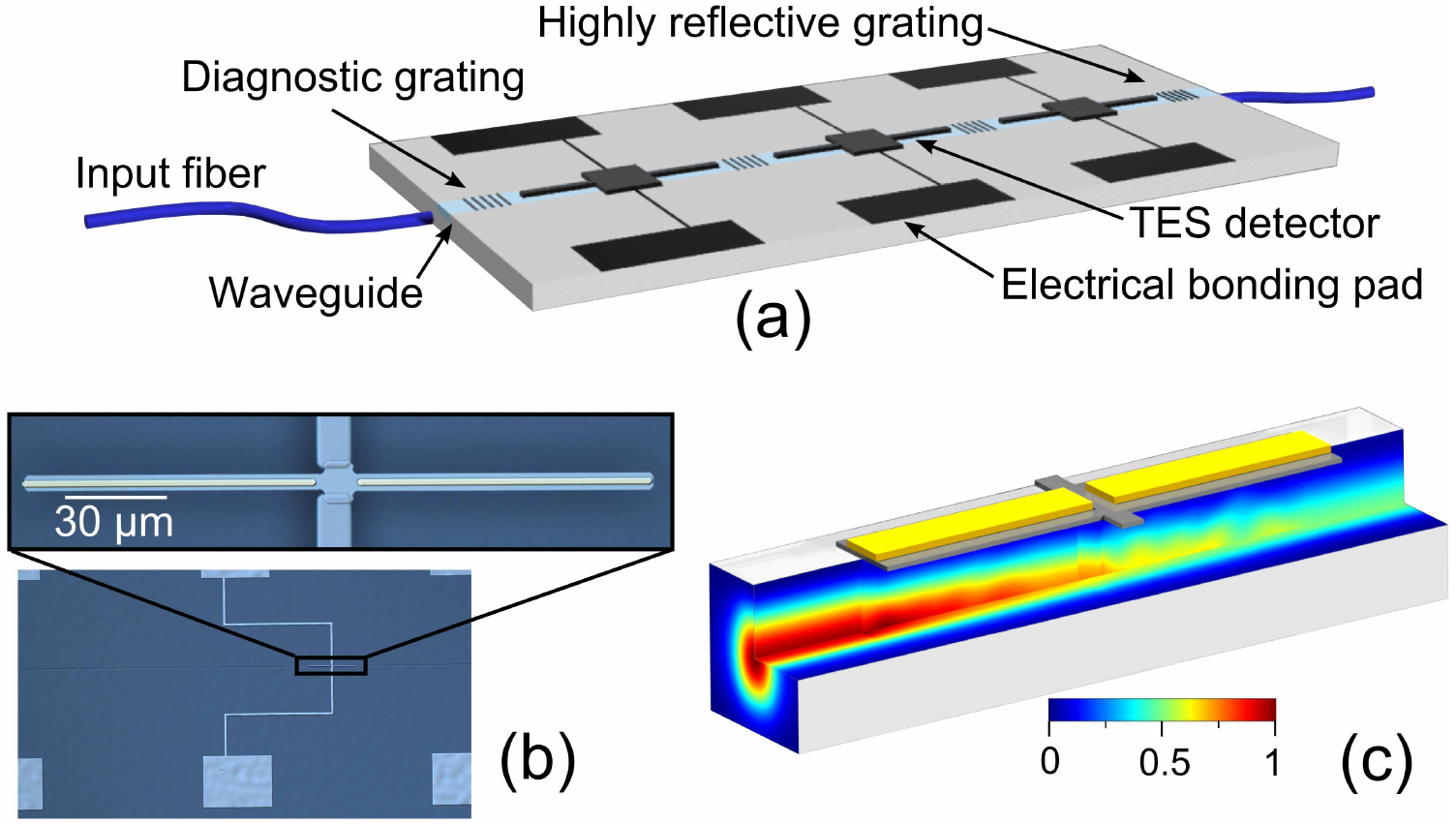}
\caption{On-chip photon detection scheme.  (a) Three TES detectors with extended absorbers are operated in series on a single waveguide, resulting in a combined 79~\% single-pass detection efficiency after correcting for fiber-coupling losses.  Integrated Bragg gratings and two-way fiber coupling allows for the precise determination of each device efficiency without additional assumptions. (b) The extended-absorber detector utilizes a gold spine to increase thermal conduction and allow absorbed energy (from photons) to be detected by the TES (square device, center).  (c) Simulated mode profile as light propagating in the waveguide is absorbed by the detector due to evanescent coupling.}
\label{concept}
\end{figure}

The concept behind the on-chip photon detectors used in this work is shown in Figure \ref{concept}.  A series of three TES photon-counting detectors is integrated with a UV-written single-mode waveguide.  In order for this detector system to operate with high efficiency, number-resolution, and low scattering and reflective losses, both the optical characteristics of the waveguide/detector system and the electro-thermal characteristics of the TES device must be engineered appropriately.  The basic designs of both systems are described in the following sections.

\subsection{Optical design}
The waveguiding component of the fabricated device is composed of a silica-on-silicon planar structure comprising a silicon substrate, thermal oxide under-cladding, and a photosensitive germano-silica core layer deposited via Flame Hydrolysis Deposition (FHD). Lateral guidance was achieved via direct UV writing into the photosensitive core layer, a process that locally increases the refractive index, resulting in a buried waveguide of Gaussian profile with an index contrast of $\sim 0.3~\%$~\cite{Svalgaard:98}. The absence of an overcladding in this design, as in previous work~\cite{Gerrits2011}, facilitates evanescent coupling of the waveguide mode into the tungsten TES at the surface.  Further details on the effectiveness of this coupling are given in Section \ref{opticalModelingSection}.

In order to obtain in-situ waveguide loss and detector absorption measurements for this device, a set of seven weak Bragg gratings were co-patterned with the waveguide via the process of Direct Grating Writing (DGW)~\cite{Emmerson:02}.  In DGW, a pair of focused continuous wave (CW) UV laser beams are overlapped to give an interference pattern. This pattern can then be modulated as it is translated along the sample to imprint either a continuous waveguide or a Bragg grating. The modulation pitch determines the Bragg wavelength and, due to the small spot size of the writing beam, wide spectral detuning is possible. 
The Bragg gratings were spectrally separated between 1500 nm and 1620 nm and were 1.5 mm in length. A Gaussian apodization profile was used to suppress grating sidebands. Additionally, a single high-reflectivity in-band grating 4 mm in length was added at the output end of the waveguide to permit a double-pass measurement to be made. This grating had a room-temperature center wavelength of 1553.3~nm with a 3~dB bandwidth of 0.6~nm.  This center wavelength shifted to 1552.0~nm when cooled to the operating temperature of the TES due to thermal contraction.

Both ends of the waveguide were pigtailed with commercial fiber V-grooves, attached using an index-matched adhesive. The device was then mounted on a silicon base plate, which provided support and thermal anchoring of the fiber pigtails.


\subsection{Detector design}

The photon detectors used in this experiment are optical TESs, based on devices used in a previous experiment \cite{Gerrits2011} but modified with additional absorbers to increase interaction device efficiency.  The TES is a calorimetric device that relies on its minuscule heat capacity (because of its small size and low temperature) and the ultra-sensitive temperature dependence of its resistance (the slope of its superconducting-to-normal transition) to allow the detection of extremely small amounts of absorbed energy -- in this case single near-infrared photons.  TESs of various types have been adopted to detect photons from mm-wave up to gamma rays, and have been made of a variety of materials such as titanium, tungsten, and molybdenum-copper bilayers \cite{IrwinHilton}. Their geometry typically consists of a small square of thin-film low-$T_c$ superconductor, often attached to a larger absorber made of a different (not necessarily superconducting) material that is sufficiently thermalized to the TES such that the absorbed energy heats the superconductor.  For the fiber-coupled tungsten optical TES a separate absorber is not necessary as the tungsten film itself is reasonably opaque in the near-IR \cite{MillerAJ03} (an opacity that can be made nearly 100~\% at a design wavelength by embedding the device in a dielectric resonant cavity \cite{LitaAE08}).  Such a device is known as a ``self-absorber" TES.  We have previously shown \cite{Gerrits2011} that this self-absorber design can be used to couple near-IR photons from a silica-on-silicon waveguide, but the efficiency was low ($7.2~\% \pm 0.5~\%$) due to the small overlap of the guided mode with the absorber and its short interaction length (only 25~$\mu$m in that case).

In this work we describe a new approach to coupling light from an on-chip waveguide into a TES detector, making use of a separate-absorber geometry that increases the interaction length with the waveguide while maintaining a low enough total heat capacity to retain the photon-number resolving capability, which distinguishes the TES from all other types of single-photon detectors.  Choosing to increase the interaction length while maintaining roughly the same extinction coefficient due to absorption in the device helps to reduce mode-mismatch at the interface between the bare waveguide and the device, which would cause unwanted reflections.  The basic concept of this new device is shown in Fig. \ref{concept} (b).  The active TES device is a $10~\mu{\rm{m}}~\times~10~\mu{\rm{m}}$ square of tungsten 40~nm thick that operates at its transition temperature of $\approx$ 84 millikelvin (mK).  Attached to the device are two 100~$\mu$m long absorbers consisting of a bottom tungsten layer 3.5~$\mu$m wide and 40~nm thick, along with a gold layer $2~\mu{\rm{m}}~\times~80~{\rm{nm}}$.  The tungsten and gold layers are deposited by DC sputtering directly on top of the written waveguide structure and patterned by optical lithography and liftoff \cite{Gerrits2011, Calkins2011}.  The majority of the evanescent photon absorption from the waveguide happens along the length of the two absorbers.  The purpose of the gold spine is to increase heat conduction so that the energy of absorbed photons will be transmitted to the TES before escaping into the substrate.  The details of the thermal model describing this device are contained in Section \ref{detectorModelingSection}.


\section{Modeling}

\subsection{Optical modeling}\label{opticalModelingSection}

The optical absorption properties of this new detector design were analyzed numerically using a fully vectorial mode-solver based on film mode matching. The modeling revealed that although the low refractive index of gold (${\it n} = 0.38$) acts to push the optical mode away from the absorbing surface, this is more than compensated for by the increased interaction length possible with a thermal gold spine. Further, a scan of possible tungsten film thickness showed that the detrimental effect of the low index of gold is minimized when the thickness is at least 40~nm. The chosen device geometry was modeled with the resulting mode profile shown in Fig.~\ref{concept}(c) -- the smooth contours highlight the minimal distortions to the optical mode caused by the metallic layers. In fact, the simulations reveal that the effective modal index varies by less than 0.1\% along the entire device, resulting in no appreciable reflective or radiative losses from the guiding region due to the detector. This is of critical importance if we wish to multiplex these detectors, since any losses between the detector and non-detector regions of the waveguide chip would contribute to a decrease in device efficiency. The excellent mode-matching and similarity in effective index between the different regions ensures that many devices can be placed one after another with no appreciable scattering or reflective losses.

The evanescent field of the guided optical mode is absorbed by the tungsten/gold tails directly above the guiding layer. The simulations predict an absorption coefficient of 32.6 cm$^{-1}$ (2.9 cm$^{-1}$) for the TM-like (TE-like) guided modes. For the whole double-fin device (total length 210~$\mu$m) including the 10~$\mu$m TES square, an eigenmode-expansion method is used to propagate the modes under the detector to arrive at a predicted device efficiency of 50.3~\% (6.1~\%) for a TM-like (TE-like) input mode. Multiplexing three detectors one after another is thus predicted to achieve a total detector efficiency of 87.7~\% for a TM-like input mode. 


\subsection{Detector modeling}\label{detectorModelingSection}

When a photon is absorbed along the tails of these detectors, that energy must be transmitted to the active TES at the center before escaping through some other thermal path.  At these temperatures the mechanism responsible for this thermal conduction is electron diffusion, and therefore its strength is proportional to both the ease with which electrons traverse through the material (electrical conductivity) and the average energy carried by each electron ($k_B T$, where $k_B$ is the Boltzmann constant).  The heat flow $\dot{Q}_{WF}$ along the length of a metal with electrical conductivity $\sigma$ and cross-sectional area $A$ is described by the Wiedemann-Franz law \cite{WiedemannFranz}:
\begin{equation}
\dot{Q}_{WF}(x) = \sigma~A~L~T_e(x)~\frac{\partial T_e(x)}{\partial x}
,\end{equation}
where $T_e$ is the electron temperature at position $x$ in the wire and $L$ is the Lorenz constant ($L = {}^{\pi^2 k_B^2}/_{3 e^2} = 24.4~{\rm{nW}}~\Omega~{\rm{K}}^{-2}$, where $e$ is the electron charge).  This conduction along the length of the spine is in competition with the thermal escape mechanism of electron-phonon coupling:
\begin{equation}
d \dot{Q}_{e{\text{--}}p}(x) = \Sigma~A~\left( T_e(x)^5 - T_p^5 \right)~d x
,\end{equation}
where $\Sigma$ is a material parameter quantifying the strength of the electron-phonon interaction and $T_e$ ($T_p$) are the electron (phonon) temperatures.  The fact that the lateral heat conduction strength is linear with electron temperature and the heat escape strength carries an exponent of 5 is very important to the design of these devices.  In the end, it means that a lower device operating temperature ($T_c$) is advantageous for thermalization of the device.  In this study, the fabrication of devices did not focus on optimizing $T_c$ for this purpose, and the relatively high $T_c$ of $\approx$ 84~mK limited the maximum length of the detectors.

The temperature evolution of the detector tails can be determined by a one-dimensional heat equation:
\begin{equation}
\gamma~A~d x~T(x,t)~\frac{d T(x,t)}{d t} = -\frac{\partial \dot{Q}_{WF}(x,t)}{\partial x} d x - d \dot{Q}_{e{\text{--}}p}(x,t)
,\end{equation}
where $\gamma$ is a material parameter characterizing the low-temperature electron specific heat of the metal (heat capacity $C = \gamma V T$).  Finally, the temperature evolution of the TES itself can be determined by use of standard models of TES electrothermal modeling \cite{IrwinHilton}.

Table \ref{tableMaterialParams} lists pertinent material parameters for tungsten and gold.  Note that the values for electrical conductivity assume bulk material, which will be valid so long as all device dimensions are much longer than the electron mean free path $l$.  The final column shows the expected thermal conductivity $\kappa$ of a thin film 80~nm thick at temperature $T = 100~{\rm{mK}}$.  The $\kappa$ value for gold is based on the assumption that the mean free path will be limited by and equal to the thickness of the film.  The mean free path is calculated as $\kappa = {}^{n e^2 l L T}/_{m v_F}$ where $n$ is the free-electron density, $m$ is the electron mass, and $v_F$ is the mean Fermi velocity.

\begin{table}[h]
\caption{Thermal and electrical material parameters}
\centerline{
\begin{tabular}{|c||c|c|c|c|c|}
\hline \\ [-1.5ex]
Material & $\gamma~[{\rm{aJ}}~\mu{\rm{m}}^{-3}~{\rm{K}}^{-2}]$ & $\Sigma~[{\rm{nW}}~\mu{\rm{m}}^{-3}~{\rm{K}}^{-5}]$ & $\sigma_{\rm{bulk}}~[\Omega^{-1}~{\rm{m}}^{-1}]$ & $l_{\rm{bulk}}~[{\rm{m}}]$ & $\kappa_{\rm{expected}}~[{\rm{W}}~{\rm{m}}^{-1}~{\rm{K}}^{-1}]$ \\ [0.5ex]
\hline \\ [-2ex]
W & $140$ \cite{Kittel} & $0.4$ \cite{Cabrera2008} & $5 \times 10^6$ \cite{note:Wconductivity} & $2.8 \times 10^{-9}$ \cite{Kittel, Feenan} & $0.0122$ \cite{Kittel, Feenan} \\
Au & $71.4$ \cite{Kittel} & $2.6$ \cite{Echternach} & $2.23 \times 10^{10}$ \cite{Ho} & $1.88 \times 10^{-5}$ \cite{Ho} & $0.232$ \cite{AshroftMermin} \\ [0.5ex]
\hline
\end{tabular}
}
\label{tableMaterialParams}
\end{table}

The advantage of using gold as a conductive spine material is seen in its nearly $20 \times$ higher thermal conductivity for this set of conditions.  In addition, its lower specific heat allows the addition of more material while maintaining similar total heat capacity.  However, the higher electron-phonon coupling strength in gold enhances the energy escape mechanism.  The right balance must be found in order to ensure a functional device.

\begin{figure}
\centering
\includegraphics[width=5.5in]{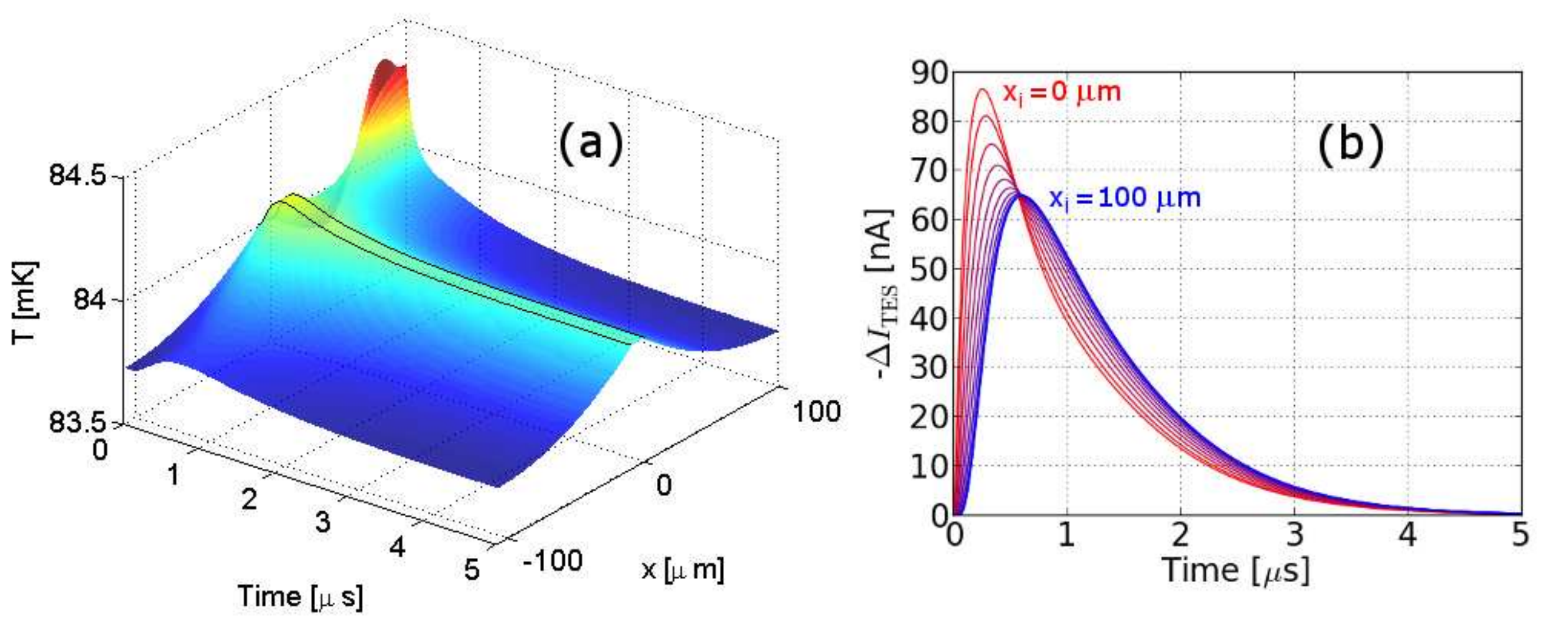}
\caption{Operation of extended-absorber device.  (a) Simulation of electron-system temperature evolution vs position on device after a 1550 nm photon is absorbed at x = +80 $\mu$m on one of the tails at $\it{t}$ = 0.  Black lines indicate borders of 10 $\mu$m square at center of device and show the temperature evolution of the TES itself.  (b) Resulting simulated change in current flowing through the device, now shown for multiple impact points between $x_i = 0 \mu$m (red) and $x_i = +100 \mu$m (blue) in 10 $\mu$m steps.}
\label{fig_simulation}
\end{figure}

In order to understand how the choice of device geometry will affect the photon-number resolution performance of the device, we used a finite-element simulation to predict the ideal current-pulse that will be measured at the TES due to photon absorption events at many locations along the length of the absorbers.  These predictions, combined with the expected noise spectrum of our TES \cite{note:TESnoise}, allows us to estimate the energy resolution of a device with a particular set of design parameters.  A sample of the thermal modeling is shown in Fig. \ref{fig_simulation} for the particular design that is the subject of this work.

\begin{figure}
\centering
\includegraphics[width=4.5in]{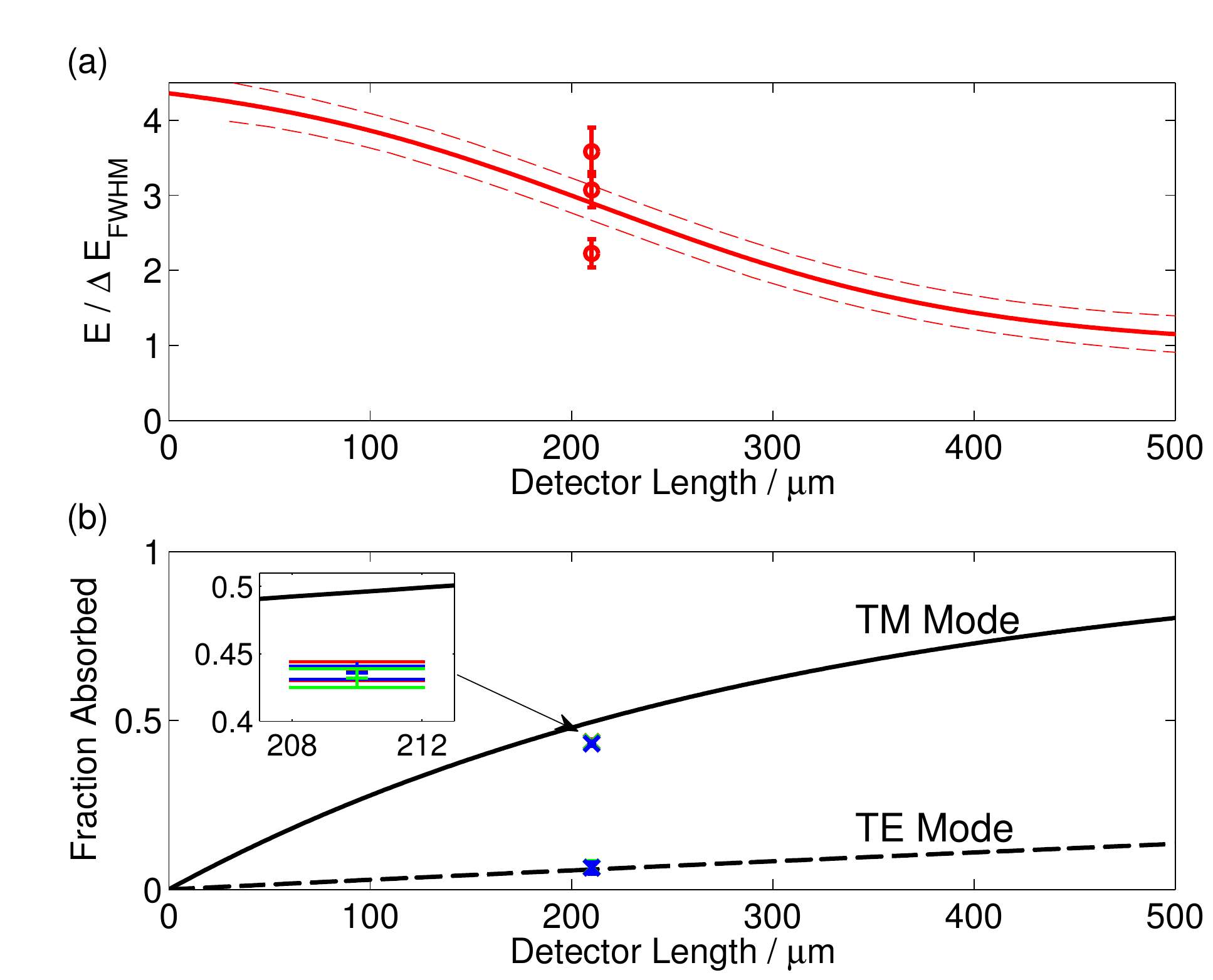}
\caption{Simulated and measured detector performance. (a) Thermal modeling. The red line shows the predicted energy resolution of the device as a function of detector length. Dotted lines indicate the standard error on the model. The circles represent the measured energy resolution of the three fabricated detectors.  The spread of measured energy resolution values is likely due to differences in the superconducting-to-normal transition properties between the three detectors. (b) Optical modeling. The black lines show the predicted optical absorption efficiency of the TM and TE modes as a function of whole detector length. The inset shows the measured photon-detection efficiency of the TM mode for the three detectors.}
\label{fig:modelingLength}
\end{figure}

This thermal modeling was used in combination with the optical modeling of Section \ref{opticalModelingSection} to search the available parameter space for a design that maximizes absorption efficiency while maintaining photon-number resolution.  This parameter space has many dimensions and several constraints due to fabrication capabilities.  Considering only a single free parameter and easily obtainable values for all other parameters, Figure \ref{fig:modelingLength} shows simulated detector efficiency and energy-resolution as a function of absorber length.  Our choice of testing the 2~$\times$~100~$\mu$m devices was determined by the energy resolution requirement.

\section{Device measurements}

In order to determine the efficiency of each photon detector with the minimum number of assumptions, several steps were taken.  First, the coupling of fiber at both ends of the waveguide under test allows a direct measurement of not only the efficiency of each detector but also the coupling loss at each end of the waveguide.  The details of this procedure are given in Section \ref{QE_section}.  Second, the inclusion of diagnostic Bragg gratings at several locations along the waveguide allows a direct measurement of not only the waveguide propagation loss but also provides an additional measurement of the fraction of light that is absorbed by each detector.  These measurements are described in Section \ref{Absorption_section}.


\subsection{Quantum efficiency measurements}\label{QE_section}

\begin{figure}
\centering
\includegraphics[width = 4.5in]{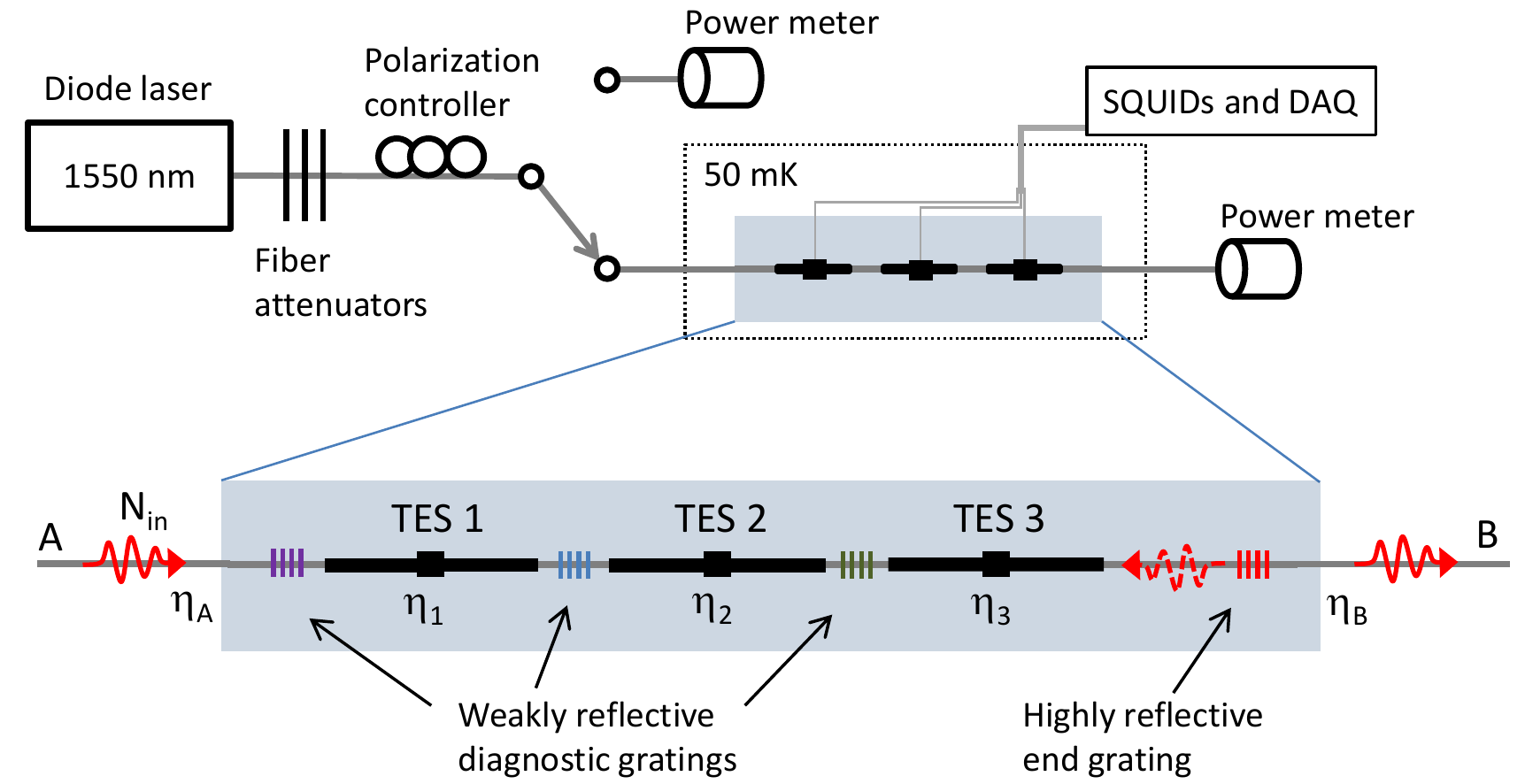}
\caption{Schematic of the experimental setup. Each TES requires its own SQUID and read-out electronics. Bottom: detailed view of the on-chip TES showing the different launch directions A and B, the three detector efficiencies, $\eta_1$, $\eta_2$, $\eta_3$ and the fiber-coupling efficiencies, $\eta_A$ and $\eta_B$. The weak reflectivity integrated Bragg gratings are used to make a room-temperature characterization of device performance whilst the high-reflectivity grating at the end of the device provides a double-pass of the detectors for an appropriately tuned source launched into port A.}
\label{setup}
\end{figure}

Measurement of the total device and individual TES detection efficiency was done following a similar scheme as described earlier~\cite{Gerrits2011}. In the present study, we measured the photon response for three in-line detectors simultaneously. Figure~\ref{setup} shows a schematic of the experimental setup. A $1550$~nm pulsed diode laser produced nanosecond pulses at a repetition rate of $100$~kHz, typically at a few $\mu$W average power. The optical fiber switch and calibrated power meter allowed both a measurement of the un-attenuated laser power and a calibration of each of the three in-line variable fiber attenuators~\cite{MillerAJ2011}. Subsequent to these measurements the laser pulses were attenuated to the single-photon regime by use of all three calibrated attenuators, and this photon flux is switched towards the device.  By this, we can determine the input mean photon number per laser pulse, $\bar{N}_{\rm in}$. The fiber polarization controller allowed tuning the polarization entering the waveguide chip. We used polarization-maintaining fiber after the optical fiber switch all the way to the waveguide chip, which was mounted inside a copper shield attached to the cold-stage of a commercial adiabatic demagnetization refrigerator (ADR) at a temperature of $\sim$~50 mK.  The detector region consisted of an array of three TESs with $100~\mu$m long absorbers on each side. A high-reflectivity grating was placed on one end of the detector array to allow double-pass detection at the grating peak wavelength of 1552.0~$\pm$~0.1~nm. When measuring the photon response of the detector array for each direction (A~$\rightarrow$~B and B~$\rightarrow$~A), an accurate determination of all individual TES detection efficiencies as well as the fiber-waveguide coupling efficiencies was possible.

Light pulses with known mean photon number $\bar{N}_{\rm in}$ can be coupled to the structure and detector array from the two facets of the device (A and B). The reflection gratings are weak reflectors at different wavelengths, ranging from $1500$~nm to $1620$~nm. These gratings are used to measure the in-situ loss of the waveguide structure, revealing the portion of loss associated with the placement of a detector on top of the waveguide structure as well as the propagation loss in the waveguide itself, as described in Section~\ref{opticalLossResultsSection}. When measuring the individual detector response for both possible propagation directions (A $\rightarrow$ B and B $\rightarrow$ A), a total of six measurements are done. The outcomes of all six measurements are captured by six equations:

\begin{eqnarray}
\label{cal_elements}
\bar{N}_1 &=& \eta_{\rm A} e^{-\alpha L_1} \eta_1\bar{N}_{\rm in}\\ \nonumber
\bar{N}'_1 &=& \eta_{\rm B} e^{-\alpha \left(L_1 + 2 L_2\right)} (1-\eta_3)(1-\eta_2)\eta_1\bar{N}_{\rm in}\\ \nonumber
\bar{N}_2 &=& \eta_{\rm A} e^{-\alpha \left(L_1 + L_2\right)} (1-\eta_1)\eta_2\bar{N}_{\rm in}\\ \nonumber
\bar{N}'_2 &=& \eta_{\rm B} e^{-\alpha \left(L_1 + L_2\right)} (1-\eta_3)\eta_2\bar{N}_{\rm in}\\ \nonumber
\bar{N}_3 &=& \eta_{\rm A} e^{-\alpha \left(L_1 + 2 L_2\right)} (1-\eta_1)(1-\eta_2)\eta_3\bar{N}_{\rm in}\\ \nonumber
\bar{N}'_3 &=& \eta_{\rm B} e^{-\alpha L_1} \eta_3\bar{N}_{\rm in}\\ \nonumber
\\ \nonumber
\end{eqnarray},
where $\bar{N}_{\rm 1/2/3}$ ($\bar{N}'_{\rm1/2/3}$) are the measured mean photon numbers at TES 1, TES 2 and TES 3, respectively for propagation direction A $\rightarrow$ B (B $\rightarrow$ A). $\bar{N}_{\rm in}$ is the derived input mean photon number inside the polarization-maintaining fiber leading towards the waveguide structure. $\eta_{\rm A/B}$ are the fiber-waveguide coupling efficiencies. $\alpha$ is the measured waveguide-propagation extinction coefficient (Section \ref{opticalLossResultsSection}), and $L_1$ ($L_2$) is the length of the waveguide between the edge of the chip and the first detector (between the first and second detector). As only five unknowns exist, the parameters are overdetermined and a simple least-mean-squares fitting routine using Eq.~\ref{cal_elements} yields the individual detection efficiencies.  To generalize this technique, the use of $N$ detectors in series results in $2 N$ measurements and $N + 2$ unknowns, and therefore $N = 3$ is the minimum number of detectors needed in order to overdetermine the resulting set of parameters. The above relations allow us to accurately extract values for both the detection efficiencies for photons that have been coupled into the waveguide and, independently, the coupling efficiencies at each facet of the device. 


\subsection{Optical loss measurements}\label{Absorption_section}

In addition to the quantum efficiency measurements conducted using the TES detectors, room-temperature characterization was conducted using the integrated Bragg gratings with high-intensity light. The grating-based loss measurement technique detailed in \cite{Rogers:10} was used to provide a relative power profile along the length of the waveguide. This ratiometric approach allows the loss profile to be determined independent of any coupling losses, and since gratings are positioned between each TES element, the loss associated with each of the three detector regions may be determined. 

Characterization was carried out both before and after the deposition of the TES layer. The former of these measurements allowed the waveguide loss to be extracted, while the latter provided absorption values for each of the detectors.  In each case, a reflection spectrum was collected while launching broadband light into each end of the waveguide. Least-mean-squares fitting of the resulting Bragg reflection peaks was then conducted to obtain values for the peak power reflected by each grating.  Section \ref{opticalLossResultsSection} details the results of these measurements.


\section{Results}

Table~\ref{tab:DEMeas} presents the individual detection efficiencies $\eta_{\rm 1/2/3}$ and coupling efficiencies $\eta_{\rm A/B}$ based on all measurements for both polarization states.  Table~\ref{tab:absorptionMeas} presents the measured detector absorption $\eta_{\rm abs,1/2/3}$ arising from the grating reflection measurements.  The agreement between most of these values ({\it i.e.}, $\eta_{\rm 1/2/3} = \eta_{\rm abs,1/2/3}$) is a demonstration of both the inherent 100~\% internal detection efficiency of the optical TES (a property that has been reported elsewhere \cite{Gerrits2011, MillerAJ2011}), as well as the lack of significant inherent scattering mechanisms between the detectors.  The few-percent mismatch for some of the values may be the result of differing surface contaminations causing additional losses in the two separate measurements.

\begin{table}
\caption{Derived individual detector efficiencies $\eta_{\rm 1/2/3}$ and fiber-waveguide coupling efficiencies $\eta_{\rm A/B}$. Derivations based on Eqn.~\ref{cal_elements}}
\label{tab:DEMeas} 
\centerline{
\begin{tabular}{|c|c|c|c|c|c|}
\hline
polarization &  $\eta_{\rm 1}$ & $\eta_{\rm 2}$ & $\eta_{\rm 3}$ & $\eta_{\rm A}$ & $\eta_{\rm B}$\\
\hline
TM & $43.7\% \pm 0.7\%$ & $43.6\% \pm 0.5\%$ & $43.2\% \pm 0.7\%$ & $22.1\% \pm 0.3\%$ &$14.8\% \pm 0.2\%$\\
TE & $6.5\% \pm 1.8\%$ & $6.6\% \pm 1.7\%$ & $6.4\% \pm 1.7\%$ & $8.1\% \pm 1.8\%$ &$8.4\% \pm 2.1\%$\\
\hline
\end{tabular}
}
\end{table}

\begin{table}
\caption{Classically obtained losses at each detector element $\eta_{\rm abs, 1/2/3}$, based on a grating based loss measurement.}
\label{tab:absorptionMeas}
\centerline{
\begin{tabular}{|c||c|c|c|c|c|}
\hline
polarization &  $\eta_{\rm abs, 1}$ & $\eta_{\rm abs, 2}$ & $\eta_{\rm abs, 3}$ \\
\hline
TM & $43.2\% \pm 0.6\%$ & $48.8\% \pm 0.6\%$ & $48.2\% \pm 0.6\%$ \\
TE & $6.3\% \pm 0.6\%$ & $4.4\% \pm 0.7\%$ & $5.3\% \pm 0.6\%$ \\
\hline
\end{tabular}
}
\end{table}


\subsection{Quantum efficiency results}
\begin{figure}
\centering
    \includegraphics[width=4in]{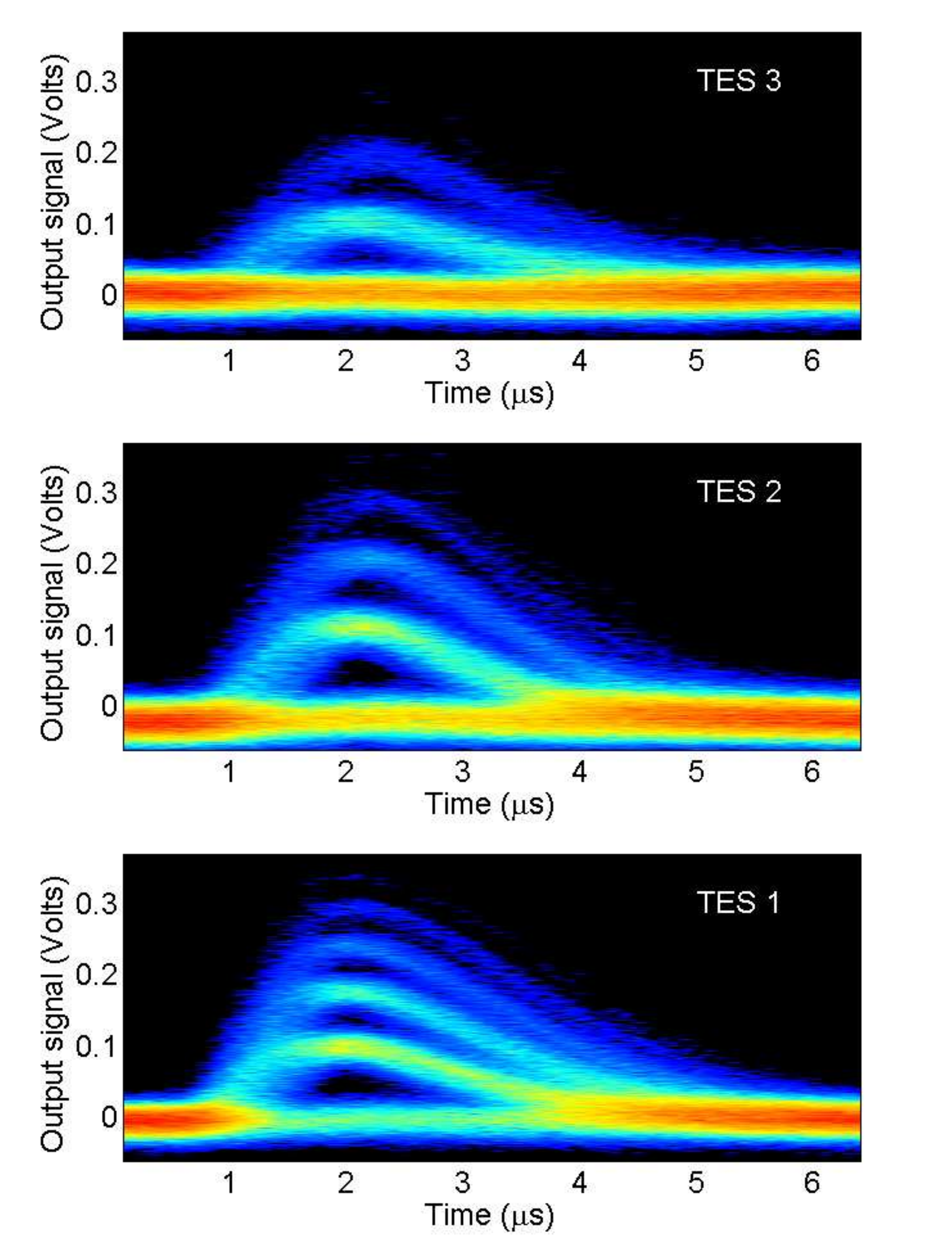}
\caption{Response curve density plot for all three TES, while sending triggered photon pulses from A to B. Pulse-collections of increasing height correspond to increasing numbers of photons detected per pulse.  The mean number of photons per pulse decreases as the light travels from detector 1 to detector 3 (bottom to top) due to absorption of the preceeding TES.}
\label{traces}
\end{figure}

We measured the detector array response for various input mean photon numbers, ranging from $\sim$~0.1 -- 3, and observed consistent efficiency results for all input states.  Figure~\ref{traces} shows an example of the response of the TES detector array, while sending photon pulses from A to B. The light is first absorbed by TES 1 (bottom) and subsequently absorbed by TES 2 (center) and TES 3 (top). The graph shows the trace densities for 8192 individual laser pulse responses, measured simultaneously from the three detectors. The photon-number distribution of the input coherent state can be clearly observed in the traces, showing the good photon-number resolution of the devices. The mean photon-number response decays as the light pulse travels from A to B, due to the absorption at each of the previous detectors.

The derived detection efficiencies for the TM input polarization are $43.7~\% \pm 0.7~\%$, $43.6~\% \pm 0.5~\%$ and $43.2~\% \pm 0.7~\%$ for TES 1, TES 2 and TES 3, respectively. The combined efficiency of all three detectors is $79~\%\pm 2~\%$ \cite{note:DEuncertainty}.  Additionally, by employing a tunable CW laser source we could utilize the high-reflectivity grating at the end of the device, resulting in a measured increase of this combined detector efficiency to $88~\%\pm 3~\%$. Since light could only be launched into port `A' for this measurement, we calculated $\eta_{1/2/3}$ simply using the measured photon counts assuming that $\eta_A$ and $\alpha$ remained unchanged. This result is consistent with the measured $\sim$~50~\% reflectivity of the terminal grating.

These measured efficiency values are the probability that a photon inside the waveguide will be detected by either of the three detectors. This does not account for the coupling efficiency from the optical fiber to the waveguide. In this case the coupling efficiencies are $22.1~\%\pm 0.3~\%$ and $14.8~\%\pm 0.2~\%$ for the A and B side, respectively. This relatively low coupling efficiency is due to misalignment of the fiber v-groove pigtail with the waveguide when cooling the system to cryogenic temperatures, as well as fiber/fiber coupling losses outside of the cryostat. We employed a different method of gluing the fiber v-groove pigtail to the waveguide chip compared to a previous study where we were able to achieve $\sim40~\%$ coupling efficiencies at the operating temperature. However, we are confident to achieve the typically high room-temperature coupling efficiencies of $>60~\%$ at cryogenic temperatures future studies.


\subsection{Optical loss results}\label{opticalLossResultsSection}

\begin{figure}[!htp]
\centering
	\includegraphics[width=0.7\textwidth]{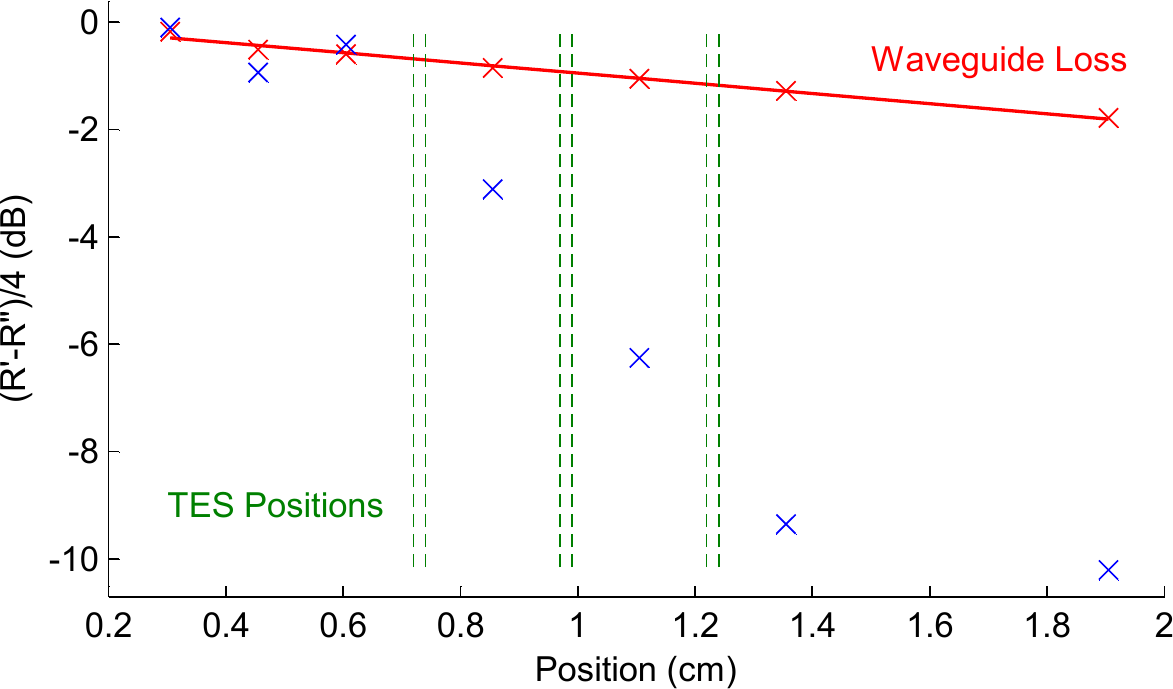}
\caption[]{The relative power profile obtained for the TM-like mode of the device, each point corresponding to a grating. Red (blue) marks indicate measurements taken before (after) TES deposition.  {\it R'} and {\it R''} are the grating peak powers measured in reflection from the forward and reverse launches, respectively.  Both waveguide propagation loss and detector absorption can be inferred from these measurements.}
\label{fig:gratingabs}
\end{figure}

Figure~\ref{fig:gratingabs} displays the relative power profile of the device for the TM polarization. The difference in relative power on either side of each of the tungsten regions, after subtracting the measured waveguide loss over this distance ($0.920 \pm 0.020$ dB/cm and $0.947 \pm 0.023$ dB/cm for TE and TM respectively), gives an estimate of the power coupled into the detector. The average absorption at each TES was thus found to be 46.7\% ($43.2\% \pm 0.6\%$, $48.8\% \pm 0.6\%$, $48.2\% \pm 0.6\%$) for TM  and 5.3\% ($6.3\% \pm 0.6\%$, $4.4\% \pm 0.7\%$, $5.3\% \pm 0.6\%$) for TE. These values include both the absorption within the tungsten and any scattering at the interfaces on either side of this region, though modeling suggests that this scattering is small. The values are in very good agreement with the detection efficiency measurements above, implying that radiative and reflective losses due to the detector are minimal, an important conclusion that could not be verified in an earlier study~\cite{Gerrits2011}. 

\section{Conclusions}

We have demonstrated a high-efficiency number-resolving photon detector integrated on-chip with a single-mode waveguide -- a critical component for many applications in quantum information.  In addition, we have shown the utility of chip-edge optical-fiber coupling and integrated Bragg gratings for device diagnostics, allowing us to determine device and coupling efficiencies to within a few percent without additional assumptions.  The fact that the two independent measurements of detector absorption and detection efficiency are consistent demonstrates the absence of additional photon-loss mechanisms, meaning that these devices (even with low efficiency) are analogous to 100~\% efficient heralding detectors.  This initial test of the extended-absorber TES demonstrates the potential for high-efficiency waveguide-integrated photon-counting detectors.  Additional improvements and optimization of this design hold the promise of achieving near-unity efficiency while retaining true photon-number resolution -- one of the ultimate goals for detectors in any quantum optical system.


\section*{Acknowledgments}
This work was supported by the NIST Quantum Information Initiative and by the EPSRC (Engineering and Physical Sciences Research Council) (grant no. GR/S82176/01), EU IP (Integrated Project) Q-ESSENCE (Quantum Interfaces, sensors, and communication based on entanglement). IAW acknowledges a Royal Society/Wolfson Research Merit Award.

\end{document}